# Antagonising explanation and revealing bias directly through sequencing and multimodal inference

Luís Arandas
INESC-TEC, Faculty of Engineering
University of Porto
luis.arandas@inesctec.pt

Mick Grierson
CCI, University Arts London
London, United Kingdom
m.grierson@arts.ac.uk

Miguel Carvalhais
i2ADS, Faculty of Fine Arts
University of Porto
mcarvalhais@fba.up.pt

## ABSTRACT

Deep generative models produce data according to a learned representation, e.g. diffusion models, through a process of approximation computing *possible* samples. Approximation can be understood as reconstruction and the large datasets used to train models as sets of records in which we represent the physical world with some data structure (photographs, audio recordings, manuscripts). During the process of reconstruction, e.g., image frames develop each timestep towards a textual input description. While moving *forward* in time, frame sets are shaped according to learned bias and their production, we argue here, can be considered as *going back in time*; not by inspiration on the *backward* diffusion process but acknowledging culture is specifically marked in the records. Futures of generative modelling, namely in film and audiovisual arts, can benefit by dealing with diffusion systems as a process to compute the future by inevitably being *tied to the past*, if acknowledging the records as to capture fields of view at a specific time, and to correlate with our own finite memory ideals. Models generating new data distributions can target video production as signal processors and by developing sequences through timelines we ourselves also go back to decade-old algorithmic and multi-track methodologies revealing the actual predictive failure of contemporary approaches to synthesis in moving image, both as relevant to composition and not explanatory.

## 1  Datasets (records) become predictive (revisiting, in part or completely)

Deep generative models rely on large datasets to be trained and iteratively update their internal representation to better predict specific outcomes [6]. Depending on modality (image, audio and text) datasets are *large*, pursuing better futures of representation, in language-guided image diffusion with at least photographic or manuscript records [18]. Using trained models in production will reveal parts of the dataset used, even fleetingly in both visual and linguistic element composition, as their sampling is the model's actual ability in resembling it (configured also in partial translation, see e.g. *style-transfer* [16]).

Datasets are therefore, not just possibly always predictive in nature but also, we argue, with a prediction *tied to the past* as a specifically cherished characteristic of learning from data and supposed contextual experience [4]. *Navigation-sampling* or *embedding-indexing* of trained generative models in time shows what can be (not necessarily always) an accurate representation of the physical world, even if stripped from contextual meaning with an anchor of time, found through similarities and, e.g., emergent visual patterns at the *specific* period of training [19]. When posing a way to revisit these records using text-to-image models at least two methodologies can be established as practical: 1) to use language (sets of text-prompts) to guide each timestep, and 2) to index specific dataset's element (sets of text-prompts), following previously established techniques such as (e.g.) neural and vector search (archeologically, *in part* and as *guidance* to display the actual records). *Jan Bot* by Bram Loogman and Pablo Núñez Palma (2018) illustrates that by generating films using found footage from the Eye's museum, as a predecessor to new possible language developments, see [17].

Image diffusion systems with language guidance provide a multimodal methodology from (at least) two separate datasets, which through inference, will have their longevity analysed [9]. Inference or predictive behavior through trained models is a methodology to configure how something mapped (inscribed in a specific data distribution or pattern) *could be* [10], in the actual implementation computing new data. Having the learned representation dictate which parameters can be scheduled and tractable through new programmatic and real-time algorithms, we introduce the concept of *virtual timelines* (synced event schedulers) as a potentially interesting candidate to reveal a model's ability in reconstruction with practical media production once again, e.g. films and video sequences through *synthesisers* with *extractive* character [3], exhibiting specific cultural and ethical marks [8]. Filmmakers and practitioners are able to appropriate and build upon methodologies of parameter automation which should be model-specific and survey new emerging practices (e.g., artificial cameras and parameterised visual abstraction [2, 11]).

## 2 Film (outputs) as both documental and experimental report (bias)

Using the mentioned procedures in order to produce a film, image diffusion with classifier guidance (e.g. OpenCLIP and LAION-5B) will produce rough indeterminism opposed to classifier-free guidance methodologies, e.g. DDIM or PLMS [9]. This is an added layer of interest given established control of the torch compute graph, always tied to class approximation, see [13] on determinism and classifier-free guidance. To pull each frame-step towards a textual prompt element of a set, means to endure a *divisive* concept objectively inscribed in the network itself (label, feature vector). Format wars of coordination encompass, e.g., processing sets of text prompts directly even if they are outputs themselves, as summarisation and templates for scripts (an interface with the *description of film shoot*). Natural language processing and topic modelling [15], as well as grammar induction and understanding [7], specific divergence on futures of image description and derived systems promote and are relevant to what has been recognised as *prompt engineering* (negative weights and concept contrast) [12].

By designing interfaces cross-compatible with different language models, embedding coordination with very simple camera transforms can result in ways to mechanise image production once again (irregularly), which itself can be target of explanation and further study (regarding bias and failure) [20]. Engaging in practical methods to sequence parameters of the diffusion process with textual guidance can already establish some valid research paths, in which case: through *variability* in: 1) frame-skip-steps and spacing, 2) three-dimensional field of view planes, 3) *language-guidance* ratios (in frame-shot composition opposed to masking), 4) manuscript embedding and automatic keyframe organisation (e.g. films *about* specific text), 5) camera angles transform point of view shot templates; acknowledging moving image diffusion purely from language (50% skip over the first frame, recursively) and the first frame from noise or a pure black pixel array. Variability and parameter establishment is to then maintain coordination between frame-by-frame diffusion, promoting: 1) a documental constitution for the ability to index specific records as influencers in the output [14], and 2) also an abstractive one, which can be developed by chaos in the established parameters[1]. It is also possible for other separate models to target a function of establishing sets of prompts themselves [22], multimodality as a joint effort is criticised in the automatic photo narrator *word.camera* project by Ross Goodwin (2016), where brief poems from captured images are generated using neural networks and automatically printed.

Video production stands on directional (according to perceptual laws) counters which process data in several dimensions. *Backward* diffusion processes create a disruption as if designed coming *back* from some visual disorder (practically, noise), but still moving *forward* in time [5], and with the development of each shot transform, language tweaks how the image *should look like* — shifting from still to moving image paradigms (e.g.) flow coherence, shot transition with semantic guidance, etc. Image diffusion models re-establish known theories of representation in moving image, as a new procedure evolving from what has been three-dimensional scene building (CGI) searching for another possible construction of worlds, see [1]. The works *Parallel I-IV* by Harun Farocki (2012-14) address the images of computer games, investigating their relationship with first person, and reality as "absent referent" constructed by algorithms[2].

## 3 Proposal and conclusion (futures)

Our proposal bears no resemblance to what explanation stands for but can help produce scientific knowledge according to pre-established practice *exposing* deep generative models' reconstruction (and its *bias*) capabilities directly in moving image with language (text-prompt) guidance by process of sequencing (event scheduling) and embedding coordination. Image and language models (guided diffusion), and learning overall, will contain in themselves parts of culture at a specific time, if made by (photographic and linguistic) *records* which are captured by physical recorders (even as a made-up simulation themselves), grounded on realism ventures with increasingly higher resolutions and bigger datasets.

We propose to target the development of a fair, reproducible (fuzzy), cross-modal and shared representation that can be used across different model (families) with variant architectures defining specific model (modal) functionalities of the whole film coordination, which is continuous and should provide semantic and formal parameter control, as a *template* (virtual timelines).

---

[1] Twisting the parameters from e.g. a spaced diffusion iteration, can result on blurry and abstract shapes as it is unable to approximate the desired output coherently, which can result in very clear color feedback in moving image.

[2] *e-flux*, Exhibitions at *Paço das Artes*, Jan. 28,- (2016), Retrieved 27-05-2023.

Specifically in production today (multimodal inference), we have to work *with* what we define as bias in a learned representation, as it is *specifically* what defines a model's ability in representing reality, practically time-stamped and used with both failure and success (*objective document, abstractive resample*), by resembling a specific dataset with (usually) pre-established and updatable structure [4]. We acknowledge there is danger in this statement when considering learning and agential systems as *purely* and blindly instrumental, as it has always been with conventional cameras, even if they became part of every material surface in every country of Europe.

Therefore, and through this simple ideology we claim that by making films, in the end, we continue with a framework which is not of explanation[3], is compositional by default, and should be treated carefully as a *feature* of what is in fact different from other generative systems which don't learn. We reached a historical time in which immediate-mode *virtual cameras* architect image translation methodologies, making up new realities tied to actual *lens* (see *image-to-text* [21]) and each specific video output is itself conditioned by whatever culture.

## ACKNOWLEDGMENTS

The research leading to these results was conducted at the UAL Creative Computing Institute (03-08/2022) and financially supported by the Portuguese Foundation for Science and Technology (FCT), through the individual research grant 2020.07619.BD and by the project "Experimentation in music in Portuguese culture: History, contexts and practices in the 20th and 21st centuries" (POCI-01-0145- FEDER-031380), co-funded by the European Union through the Operational Program Competitiveness and Internationalisation, in its ERDF component, and by national funds, through the Portuguese FCT.

---

[3] We acknowledge the role of journalism and ethnographic work which does indeed share production methodologies with practice raised in this manuscript.